%% file: main.tex
\documentclass[sigconf,authorversion,natbib=true,nonacm]{acmart}

\AtBeginDocument{%
  }

\usepackage{multirow}
\usepackage{amsmath}
\usepackage{relsize}
\usepackage{graphicx}
\usepackage{listings}
\usepackage{xcolor}
\usepackage{booktabs}
\usepackage{fancyref}
\usepackage{todonotes}
\usepackage{import}
\usepackage{dsfont}
\usepackage{lineno}
\usepackage{float}
\usepackage{caption}
\usepackage[skip=10pt]{subcaption}
\usepackage{bm}
\usepackage{bbm}
\usepackage{commath}
\usepackage[section]{placeins}
\usepackage{fancyhdr}
\usepackage{natbib}
\usepackage{datetime}

\DeclareMathOperator*{\argmax}{argmax}

\def\RR{\mathbb{R}}

\usepackage{algorithmic}
\usepackage[ruled,vlined]{algorithm2e}
\usepackage{makecell}
\usepackage{ifthen}
\usepackage{makecell}
\usepackage{siunitx}

\usepackage{amsthm}

\usepackage{varwidth}
\usepackage{capt-of}

\usepackage{tikz}
\usetikzlibrary{calc,matrix,chains,positioning,decorations.pathreplacing,arrows}

\newcommand{\myref}[2]{\hyperref[#2]{#1 \ref*{#2}}}

\newcommand{\na}{\textcolor{gray}{\footnotesize N/A}}

\theoremstyle{definition} 
\begin{document}

\title[NeuralFactors: A Novel Factor Learning Approach to Generative Modeling of Equities]{NeuralFactors: \\A Novel Factor Learning Approach to Generative Modeling of Equities}

\author{Achintya Gopal}
\affiliation{%
  \institution{Bloomberg}
  \streetaddress{731 Lexington Ave}
  \city{New York}
  \country{USA}}
\email{agopal6@bloomberg.net}



\begin{abstract}
The use of machine learning for statistical modeling (and thus, generative modeling) has grown in popularity with the proliferation of {time series models}, text-to-image models, and especially large language models. 
Fundamentally, the goal of classical factor modeling is statistical modeling of stock returns, and in this work, we explore using deep generative modeling to enhance classical factor models.
Prior work has explored the use of deep generative models in order to model hundreds of stocks, leading to accurate risk forecasting and alpha portfolio construction; however, that specific model does not allow for easy factor modeling interpretation in that the factor exposures cannot be deduced. In this work, we introduce \textit{NeuralFactors}, a novel machine-learning based approach to factor analysis where a neural network outputs factor exposures and factor returns, trained using the same methodology as variational autoencoders. We show that this model outperforms prior approaches both in terms of log-likelihood performance and computational efficiency. Further, we show that this method is competitive to prior work in generating realistic synthetic data, covariance estimation, risk analysis (e.g., value at risk, or VaR, of portfolios), and portfolio optimization. Finally, due to the connection to classical factor analysis, we analyze how the factors our model learns cluster together and show that the factor exposures could be used for embedding stocks.
\end{abstract}

\maketitle

\section{Introduction}

Understanding alpha and risk is a crucial component of financial modeling on equities; one way to approach modeling these elements is through statistical modeling, which further allows one to generate synthetic data.
One approach to statistical modeling of stock returns is factor modeling, i.e., modeling each stock return as the sum of a linear combination of factor returns and an idiosyncratic component. Three popular ways to build factor models are to define factor returns, e.g., Fama-French \citep{FamaFrench2015return}, to define factor exposures, e.g., Barra \citep{Barra}, or to infer factor exposures statistically, e.g., Probabilistic PCA (PPCA) \citep{ghojogh2021factor}.
In this work, we focus on a deep learning approach to learning factors exposures, which allows us to not only have a generative model over hundreds of stocks, but also allows us to utilize classical techniques for risk forecasting and portfolio construction.

Deep generative models are simply statistical models, and thus, we use deep probabilistic models for the task of modeling distributions of returns. Prior work has shown this approach to risk forecasting is competitive with classical approaches \citep{tepelyan2023generative}.

In terms of deep probabilistic modeling of financial time series, prior work has applied deep learning approaches to model a single financial time series (e.g., \cite{Buhler2020Generate, wiese2019quantgan}), or multivariate time series (e.g.,  \cite{Wiese2021Multi,Sun2023GAN,Masi2023CorrelatedTimeSeries,tepelyan2023generative, Ramirez2023ML}).
Of this work, \citet{tepelyan2023generative} (BDG\footnote{For a shorthand, we will refer to the model from \citet{tepelyan2023generative} as the Baseline Deep Generator (BDG).}) were the first to scale up to hundreds of stocks through combining machine learning with factor modeling, specifically Fama-French factor modeling. However, this approach requires choosing a set of factors to explain the correlations among stocks, and thus, if there are any undiscovered factors, this approach will not be able to discover them. In this paper, we remove this requirement of choosing a set of factors a priori and instead, our machine learning model discovers factors from scratch.

Specifically, our approach uses variational autoencoders (VAEs, \myref{Section}{sec:ciwae}), where the latent space represents the market factors and the decoder combines these with learned factor exposures (\myref{Section}{sec:methodology}). A high level diagram of our modeling methodology can be seen in \myref{Figure}{fig:final_arch}. Since we can directly interpret our model as learning factors and factor exposures, we name our model \textit{NeuralFactors}.

Factor analysis has been described as ``one of the simplest and most fundamental generative models'' \citep{ghojogh2021factor}. Through this lens of viewing factor learning as a generative modeling task, we develop a novel machine-learning based factor analysis methodology and architecture which:
\begin{itemize}
    \item generates hundreds of stock returns, outperforming prior work, namely BDG and PPCA,
    \item efficiently predicts mean and covariance, and
    \item learns risk factors and idiosyncratic alpha in an end-to-end fashion.
\end{itemize} 
We empirically illustrate the efficacy of NeuralFactors on the (point-in-time) constituents of the S\&P 500 (\myref{Section}{sec:eval}). We are able to outperform BDG and PPCA in terms of negative log-likelihood (\myref{Section}{sec:nll}) and covariance forecasting (\myref{Section}{sec:covariance}). We can apply NeuralFactors to VaR analysis; while it outperforms PPCA and BDG on the test set, GARCH still outperforms NeuralFactors in terms of calibration error (\myref{Section}{sec:var}).
Finally, our model creates portfolios that outperforms the market (\myref{Section}{sec:portfolio}). Since we can interpret our model's outputs as factor exposures, we qualitatively analyze the factors discovered by our model (\myref{Section}{sec:tsne}).

\section{Background}

The crux of our methodology is modeling conditional distributions $p(\mathbf{y}|\mathbf{x})$. To do so, we use a conditional importance-weighted autoencoder (CIWAE  \citep{burda2015iwae,ESGImp2020}). CIWAEs allow us to approximate log-likelihoods, a task which is both useful for statistical evaluation and for training using maximum likelihood estimation; further, we can use its generative capabilities to evaluate the quality of the synthetic data. 
At a high level, CIWAE is a latent variable model and, in the context of generative modeling of equities, we can interpret the latent variable as factor returns.

\subsection{Student's T Distribution}\label{sec:nig}

The density function of a Student's T distribution is\footnote{$t$ (without a subscript) will refer to time; $t_\nu$ (with a subscript) will refer to the Student's T distribution}:
$$ p(x\ |\ {t}_\nu(\mu, \sigma)) = \frac{\Gamma\left(\frac{\nu + 1}{2} \right)}{\sqrt{\pi \nu} \sigma \Gamma\left( \frac{\nu}{2} \right)} \left(1 + \frac{(x - \mu)^2}{\sigma^2 \nu} \right)^{-\frac{\nu + 1}{2}} $$
where $\Gamma$ is the Gamma function, $x \in \RR$, $\mu \in \RR$ (the mean parameter),  $\sigma > 0$ (the scale parameter), and $\nu > 4$ (the degrees of freedom where the constraint is in order to have a finite kurtosis). 

When referring to multivariate Student's T distribution $\mathbf{t}_{\bm\nu}(\bm{\mu}, \bm{\sigma})$, we will simply be referring to a product distribution:
$$ p(\mathbf{x}\ |\ \mathbf{t}_{\bm\nu}(\bm{\mu}, \bm{\sigma})) = \prod_{i=1}^{N} p(x_i\ |\ {t}_{\nu_i}(\mu_i, \sigma_i)) $$
where $\mathbf{x}$, $\bm{\mu}$, $\bm{\sigma}$, and $\bm{\nu}$ are all $N$-dimensional vectors.

\subsection{VAE and Conditional VAE (CVAE)}\label{sec:vae}

Suppose that we wish to formulate a joint distribution on an $n$-dimensional real vector $x$. For a VAE-based approach, the generative process is defined as:
$$ \mathbf{x} \sim p(\mathbf{x} | \mathbf{z}) \qquad \mathbf{z} \sim p(\mathbf{z}) $$
where $\mathbf{x}$ is the observed data and $\mathbf{z}$ is a latent variable.
VAEs use a neural network to parameterize the distribution $p(\mathbf{x} | \mathbf{z})$. 

Say we are modeling $p(\textbf{y}|\textbf{x})$, we can change the generative process to:
$$ \mathbf{y} \sim p(\mathbf{y} | \mathbf{z}, \mathbf{x}) \qquad \mathbf{z} \sim p(\mathbf{z} | \mathbf{x}) $$
For the generative process, we can see sampling from a VAE requires \textit{two sampling steps}: sampling from $p(\mathbf{z} | \mathbf{x})$ and then sampling from $p(\mathbf{y} | \mathbf{z}, \mathbf{x})$. Often $p(\mathbf{y} | \mathbf{z}, \mathbf{x})$ is referred to as the \textit{decoder}.

When fitting distributions to data, the most common method is maximum likelihood estimation (MLE):
$$ \argmax_{\theta} \sum_{i=1}^{N} \log p(\mathbf{y_i} | \mathbf{x_i}, \theta) $$
where $\{\mathbf{x_i},\mathbf{y_i}\}_{i=1}^{N}$ denotes the data we want to fit our model on. For a latent variable model, the log-likelihood is:
 $$ \log p(\mathbf{y_i} | \mathbf{x_i}, \theta) = \log \int p(\mathbf{y_i} | \mathbf{x_i}, \mathbf{z}, \theta) p(\mathbf{z} | \mathbf{x_i}) dz$$
This integral is the reason we cannot directly perform gradient descent of the total log-likelihood; VAEs instead use variational inference \citep{kingma2013vae}, i.e., optimize the evidence lower bound (ELBo). 
Since we are modeling the conditional distribution, we give the conditional version of the VAE loss:
\begin{align}
 \log p(\mathbf{y_i} | \mathbf{x_i}, \theta) &\geq \mathbb{E}_{\mathbf{z} \sim q(\mathbf{z} | \mathbf{x_i}, \mathbf{y_i})}\left[\log  \frac{p(\mathbf{y_i} |\mathbf{x_i}, \mathbf{z}, \theta)\ p(\mathbf{z} | \mathbf{x_i}, \theta) }{q(\mathbf{z} | \mathbf{y_i}, \mathbf{x_i})} \right] \label{eqn:cvae}
 \\ &= -\mathcal{L}_{\text{CVAE}}(\mathbf{y_i}, \mathbf{x_i}; \theta, \phi) 
\end{align}
using Jensen's inequality.

\subsection{CIWAE}\label{sec:ciwae}

A simple trick to help close the gap in the lower bound is to use importance weighted autoencoders (IWAE). 
Similar to CVAE, we give the conditional version of the IWAE loss:
\begin{align}
 \log p(\mathbf{y_i} | \mathbf{x_i}, \theta) &\geq \mathbb{E}_{\mathbf{z_1}, \dots, \mathbf{z_k} \sim q(\mathbf{z} | \mathbf{x_i}, \mathbf{y_i})}\left[\log  \sum_{j=1}^{k}  \frac{p(\mathbf{y_i} |\mathbf{x_i}, \mathbf{z_j}, \theta)p(\mathbf{z_j} | \mathbf{x_i}, \theta) }{q(\mathbf{z_j} | \mathbf{y_i}, \mathbf{x_i})} \right] \label{eqn:ciwae}
 \\ &= -\mathcal{L}_{\text{CIWAE}, k}(\mathbf{y_i}, \mathbf{x_i}; \theta, \phi) 
\end{align}
\citet{burda2015iwae} showed that increasing $k$ reduces the bias of the estimator.

\section{Methodology}\label{sec:methodology}

The goal of our model is to learn:
$$ p(\mathbf{r}_{t+1} | \mathcal{F}_t) $$
or, in other words, the joint distribution of $N_{t + 1}$ returns $\mathbf{r}_{t+1} \in \RR^{N_{t+1}}$ at time $t + 1$ given historical data $\mathcal{F}_t$. We use $\mathcal{F}_t$ to represent all the information available up until and including time $t$; the specific information from the set used in our model is defined in \myref{Section}{sec:features}.

In \myref{Section}{sec:problem}, we discuss how we formulate the problem as a latent variable model where the latent variable represents factor returns. In \myref{Section}{sec:decoder}, we describe our decoder model, which is parameterized in order to give linear factor exposures. Since we use a latent variable model, we use the CIWAE loss (\myref{Equation}{eqn:ciwae}), and so, in \myref{Section}{sec:encoder}, we describe how we parametrize the encoder; as opposed to prior work, our encoder does not include any encoder-specific learnable parameters. Putting all the pieces together (\myref{Section}{sec:arch_and_opt}), we describe the architecture and how we optimize our model. Finally, in \myref{Section}{sec:usage}, we explain how to use NeuralFactors for generating synthetic data as well as risk forecasting.

\input{final_arch}

\subsection{Problem Formulation}\label{sec:problem}

To model a variable number of securities, we use the following generative process:
\begin{align}
    \mathbf{z_{t+1}} &\sim \mathbf{t}_{\bm\nu_z}(\bm\mu_z, \bm\sigma_z)
\\  r_{i, t+1} &\sim p( r_{i, t+1} | \mathbf{z_{t+1}}, \mathcal{F}_t) \qquad \text{for } i=1 \text{ to } N_{t + 1}
\end{align}
where $\mathbf{z_{t+1}} \in \RR^F$ ($F$ refers to the number of latent factors).
Intuitively, $\mathbf{z_{t + 1}}$ represent the market factors for time $t+1$, i.e., $\mathbf{z_{t + 1}}$ captures all the dependence (e.g., correlations) we observe in stock returns. Using this generative process, we write the likelihood as:
$$\log p(\mathbf{r}_{t+1} | \mathcal{F}_t) = \int \left(\prod_{i=1}^{N_{T+1}}  p(r_{i,t+1} | \mathbf{z_{t+1}}, \mathcal{F}_t) \right)\ p( \mathbf{z_{t+1}})\ d\mathbf{z_{t+1}}  $$
Note that the likelihood is similar to prior work \citep{tepelyan2023generative}, where, instead of a chosen set of factors $\mathbf{F_{t+1}}$ to capture all the dependence across stock, the above formulation treats the factors as latent variables.

To perform maximum likelihood estimation (MLE) given this likelihood, we use variational inference:
\begin{align*}
    \log p(\mathbf{r}_{t+1} &| \mathcal{F}_t) \geq \mathbb{E}_{\mathbf{z_{t+1}} \sim q(\mathbf{z_{t+1}} | \mathbf{r}_{t+1}, \mathcal{F}_t)} \big[
\\ &  \sum_{i=1}^{N_{t+1}} \log p(r_{i, t+1} | \mathbf{z_{t+1}},  \mathcal{F}_t) + p(\mathbf{z_{t+1}}) - q(\mathbf{z_{t+1}}) \big]
\end{align*}

\subsection{Linear Decoder}\label{sec:decoder}

Inspired by classical factor modeling, we set:
$$p(r_{i, t+1} | \mathbf{z_{t+1}},  \mathcal{F}_t)  = p\left(r_{i, t+1}\ |\ t_{\nu_{i,t}}\left(\alpha_{i,t} + \bm\beta_{i,t}^T \mathbf{z_{t+1}},\ \sigma_{i,t} \right) \right) $$
where $\alpha_{i,t}$, $\bm\beta_{i,t}$, $\sigma_{i,t}$, and $\nu_{i,t}$ are the outputs of functions of $\mathcal{F}_t$.
In other words, the mean of the stock returns is a linear function of the factor returns ($\mathbf{z_{t+1}}$), where the factor exposures ($\bm\beta_i$) are functions of the past, and the scale ($\sigma_i$) and degrees of freedom ($\nu_i$) are functions of the past and independent of the factor returns. Note that the expected returns of a stock is a function of both $\alpha_{i,t}$ and $\bm\mu_{z}$; however, given the generative process we have defined, we can set $\bm\mu_{z}=0$ without reducing the expressivity of the model.

\subsection{Approximating the Encoder}\label{sec:encoder}

Given this simplification of the decoder $p(r_{i, t+1} | \mathbf{z_{t+1}},  \mathcal{F}_t)$, we can derive an approximation for $q(\mathbf{z_{t+1}} | \mathbf{r}_{t+1}, \mathcal{F}_t)$. While we use Student's T distribution for our prior and decoder, we approximate the posterior by approximating the Student's T distributions with a Normal distribution using moment matching.

Say the prior distribution is a Normal distribution ($\mathcal{N}\left(\bm\mu_{z}, \Sigma_z\right)$ where $\Sigma_z \in \RR^{F \times F}$ is a covariance matrix) and the decoder is a Normal distribution ($\mathcal{N}\left( \ \bm\beta_{i}^T \mathbf{z},\ \sigma^2_{i} \right)$). Note, for simplicity, we removed the time indices. In this case, we can derive the posterior $p(\mathbf{z} | \mathbf{r}, \mathcal{F})$ in closed form (note that the returns $\mathbf{r}$ are one-day ahead returns):
\begin{align}
    q(\mathbf{z} \big| \mathbf{r}, \mathcal{F}) &= p(\mathbf{z} | \mathbf{r}, \mathcal{F}) \qquad \text{(Exact posterior)}
\\ &= p\left(\mathbf{z}\ |\ \mathcal{N}\left( \Sigma_{z | B}\ (\Sigma_z^{-1} \bm\mu_z + B^T \Sigma_{x}^{-1} (\mathbf{r} - \bm{\alpha}))\ ,\ \Sigma_{z | B} \right)\right) \label{eqn:posterior}
\\ &= p\left(\mathbf{z}\ |\ \mathcal{N}\left( \mu_{z|B, r}\ ,\ \Sigma_{z | B} \right)\right)
\end{align}
where $\Sigma_{z | B} = \left(\Sigma_{z}^{-1} + B^T \Sigma_{x}^{-1} B\right)^{-1}$, $\Sigma_{x}$ refers to a diagonal matrix with $\Sigma_{x, ii} = \sigma^2_{i}$,  $\bm{\alpha}$ is a vector comprised of all the $\alpha_i$, and $B$ is a matrix comprised of all the vectors $\bm\beta_{i}$. 

Note that the mean of this posterior is similar to an L2-regularized weighted linear regression, where $B$ plays the role of features, $\mathbf{r}$ is the target,  $\Sigma_{x}^{-1}$ is the weights, and $\Sigma_{z}^{-1}$ controls the regularization. 

While \myref{Equation}{eqn:posterior} is an approximation of the posterior, we can use the importance-weighted autoencoder loss to reduce the bias introduced by this approximation (\myref{Section}{sec:ciwae}).

\subsection{Features}\label{sec:features}

Since the goal of the model is to learn factor exposures, for our feature set, we include features that would lead to classical factors: history of stock's returns, which could be used to create a momentum factor; company industry, since there tends be industry specific correlations; and company financials listed in \myref{Table}{tab:cofi}, which could be used to create a value and size factor.
Further, to allow comparison against BDG and to give the model a sense of the ``state of the world'', we give factor indices (\myref{Table}{tab:factors}, S\&P GICS Level 2, 3, and 4 indices). Finally, in order to experiment with the ability of our modeling methodology in discovering atypical factors, we include volume features (log volume traded per day) and option features (the ratio of open interest for puts and calls). 

We will refer to stock returns as $r_{i, t}$, $\mathbf{X^{ts}_{i, t}}$ to time-varying features (the factor indices, company financials, volume features, and options features), and $\mathbf{X^{static}_{t}}$ to both the time series features and static features (company industry) at time $t$.

\begin{table}[!bt]
  \centering
  \smaller
  \scalebox{0.95}{
  \begin{tabular}{ll}
\toprule
\textit{Balance Sheet} & \textit{Income Statement}\\
\midrule
DEBT\_TO\_MKT\_CAP & PE\_RATIO \\
MKT\_CAP\_TO\_ASSETS & TOT\_MKT\_VAL\_TO\_EBITDA \\
PX\_TO\_BOOK\_RATIO & COGS\_TO\_NET\_SALES \\
WORKING\_CAPITAL\_TO\_SALES & CASH\_DVD\_COVERAGE \\
TOT\_DEBT\_TO\_COM\_EQY & RETENTION\_RATIO \\
TOT\_DEBT\_TO\_TOT\_ASSET & GROSS\_MARGIN \\
INVENT\_TURN & INT\_EXP\_TO\_NET\_SALES \\
ASSET\_TURNOVER & RETURN\_COM\_EQY \\
ACCOUNTS\_PAYABLE\_TURNOVER & RETURN\_ON\_ASSET \\
ACCT\_RCV\_TURN & \\
CASH\_RATIO & \\
\toprule
\textit{Cash Flow} \\
\midrule
FREE\_CASH\_FLOW\_YIELD  \\
FREE\_CASH\_FLOW\_MARGIN  \\
CASH\_FLOW\_TO\_NET\_INC  \\
CASH\_FLOW\_TO\_TOT\_LIAB  \\
FCF\_TO\_TOTAL\_DEBT  \\
\end{tabular}
}
\caption{List of company financials features used in our model.}
\label{tab:cofi}
\end{table}

\begin{table}[!bt]
  \centering
  \smaller
  \scalebox{0.95}{
  \begin{tabular}{ccccc}
\toprule
\textit{Baseline} & \textit{Style}  & \textit{Bond} & \textit{Commodities} & \textit{International}\\
\midrule
VIX & M2US000\$ & LBUSTRUU & BCOMINTR & MXJP \\
SKEW &M1USQU &  LUACTRUU &  BCOMAGTR & MXPCJ\\
MOVE & RAV &SPUHYBDT  &  BCOMGCTR & MXGB \\ 
RAY &  RAG & & MXCN & MXCA \\
FARBAST & DJDVY & & BCOMNGTR & MXMX \\ 
REIT & MLCP& & BCOMSITR & MXEF  \\
 & MMCP& & BCOMINTR & MXEUG \\
 & MSCP & & & MXBRIC  \\
\end{tabular}
}
\caption{List of indices we used as features in our model \citep{tepelyan2023generative}.}
\label{tab:factors}
\end{table}

\subsection{Architecture and Optimization}\label{sec:arch_and_opt}

\begin{figure}
\begin{center}
\scalebox{0.9}{
\begin{tikzpicture}[
  every neuron/.style={
    circle,
    minimum size=0.3cm,
    thick
  },
  every data/.style={
    rectangle,
    minimum size=0.4cm,
    thick
  },
]

  \node [align=center,data 1/.try, minimum width=1.5cm] (input-x)  at ($ (2.5, 0) $) {$\mathbf{{X}^{static}_{t}}$};

  \node [align=center,data 1/.try, minimum width=1.5cm] (input-ft)  at ($(input-x) + (-2.0,0)$) {$r_{i,t}\ ||\ \mathbf{X^{ts}_{i, t}}$};
  \node [align=center,data 1/.try, minimum width=1.5cm] (input-fdots)  at ($(input-ft) + (-1.5,0)$) {$\cdots$};
  \node [align=center,data 1/.try, minimum width=1.5cm] (input-f1)  at ($(input-fdots) + (-1.5,0)$) {$r_{i,t-l}\ ||\ \mathbf{X^{ts}_{i, t-l}}$};

  \node [align=center,data 1/.try, minimum width=4.0cm, draw] (lstm)  at ($(input-fdots) + (-0.0,1.)$) {Sequence Model};

  \node [align=center,data 1/.try, minimum width=1.5cm,draw] (mlp-4)  at ($(input-x) + (0.0,1.)$) {MLP};

    \draw [black,solid,->] ($(input-f1.north)$) -- ($(input-f1.north) + (0,0.4)$);
    \draw [black,solid,->] ($(input-ft.north)$) -- ($(input-ft.north) + (0,0.4)$);
  \draw [black,solid,->] ($(lstm.east)$) -- ($(mlp-4.west)$) node [midway,below] {$\mathbf{h}_{2, i}$};
  \draw [black,solid,->] ($(input-x.north)$) -- ($(mlp-4.south)$);

\node [align=center,every data/.try, data 1/.try, minimum width=0.3cm] (alpha-1)  at ($ (mlp-4) + (2.0, 0.75) $) {$\alpha_{i,t}$};
\node [align=center,every data/.try, data 1/.try, minimum width=0.3cm] (beta-1)  at ($ (mlp-4) + (2.0, 0.25) $) {$\bm\beta_{i,t}$};
\node [align=center,every data/.try, data 1/.try, minimum width=0.3cm] (sigma-1)  at ($ (mlp-4) + (2.0, -0.25) $) {$\sigma_{i,t}$};
\node [align=center,every data/.try, data 1/.try, minimum width=0.3cm] (nu-1)  at ($ (mlp-4) + (2.0, -0.75) $) {$\nu_{i,t}$};

  \draw [black,solid,->] ($(mlp-4.east)$) -- ($(alpha-1.west)$);
  \draw [black,solid,->] ($(mlp-4.east)$) -- ($(beta-1.west)$);
  \draw [black,solid,->] ($(mlp-4.east)$) -- ($(sigma-1.west)$);
  \draw [black,solid,->] ($(mlp-4.east)$) -- ($(nu-1.west)$);

\end{tikzpicture}
}
\end{center}
\caption{A diagrammatic representation of stock embedder. $l$ refers to the lookback window size and $||$ denotes concatenation.}\label{fig:stock_embedder}
\end{figure}

We show the high-level diagram of our modeling approach in \myref{Figure}{fig:final_arch}. For simplicity, we will refer to the model that outputs $\alpha_{i,t}$, $\bm\beta_{i,t}$, $\sigma_{i,t}$, and $\nu_{i,t}$ as ``Stock Embedder''. The diagram of the architecture of Stock Embedder can be found in \myref{Figure}{fig:stock_embedder}.

For our stock embedder, we condition on the previous $l$ days of stock returns and time series features $\{r_{i, u}\ ||\ \mathbf{X^{ts}_{i, u}}\}_{u=t-l}^{t}$ (where $||$ denotes concatenation) using a sequence model such as an LSTM \citep{hochreiter1997long} or attention \citep{vaswani2017attention}. In our ablation studies (\myref{Section}{sec:ablation}), we experiment with different values of $l$ and the choice of LSTM vs attention.

We pass these into a one-layer multi-layer perceptron (MLP) $\{\mathbf{h}_{1, i, t}\}_{t=1}^{T}$. 
which is then passed into two layers of a sequence model giving $\mathbf{h}_{2, t}$. In both the case of LSTM and attention, we use the last hidden state. Intuitively, $\mathbf{h}_{2, t}$ summarizes the time-series information into a vector. 
$\mathbf{h}_{2, i, t}$ is concatenated with $\mathbf{{X}^{static}_{t}}$ and is passed through another two-layer MLP which outputs $\mathbf{h}_{3, i,t}$. Finally,
\begin{align*}
 \alpha_{i, t} &= \mathbf{w}_{\alpha}^T \mathbf{h}_{3, i,t} \qquad &&\bm\beta_{i, t} = W_{\beta} \mathbf{h}_{3, i,t}
\\ \sigma_{i, t} &= S(\mathbf{w}_{\sigma}^T \mathbf{h}_{3, i,t})
\qquad &&\nu_{i, t} = S(\mathbf{w}_{\nu}^T \mathbf{h}_{3, i,t}) + 4
\end{align*}
where to ensure positivity, we use softplus ($S=\log(1 + e^x)$).

To parameterize our prior, we simply using a time homogenous multivariate Student's T distribution $\mathcal{T}(\bm\nu_z, \bm\mu_z, \bm\sigma_z)$ where, similar to the stock embedder, we use softplus to ensure the scale is positive and that the degrees of freedom is greater than 4.

We then use \myref{Equation}{eqn:posterior} to compute the variational distribution $q(\mathbf{z})$ where for $\Sigma_z$ and $\sigma_x$, we compute the variances of the prior and $\sigma_{i,t}$, respectively. Using samples from $q(\mathbf{z})$, we compute the CIWAE loss (\myref{Equation}{eqn:ciwae}).

All our layers used a hidden size of 256 and dropout of 0.25. We hyperparameter tune the number of factors (\myref{Section}{sec:ablation}) and trained our model for 100,000 gradient updates using the Adam optimizer \citep{Adam2015} with a learning rate of 1e-4, weight decay (L2 regularization) of 1e-6, and batch size of 1 with IWAE loss with $k=20$.  For further stability, we use Polyak averaging \citep{Polyak1992} starting from 50,000 steps in. We compute the validation loss every 1,000 steps and, for evaluation, use the model with the lowest validation loss. Note, when we say a batch size of 1, we refer to one row of data as all the stocks of a single day.
We implemented our model in Pytorch \citep{pytorch} using Pytorch Lightning for training \citep{Falcon_PyTorch_Lightning_2019}.

For inference, we do not need to use \myref{Equation}{eqn:posterior} and simply sample from the prior distribution $\mathcal{T}(\bm\nu_z, \bm\mu_z, \bm\sigma_z)$.

\subsubsection{Time Complexity}

Since a single batch of data requires all the stocks of a single day, we find it useful to analyze the time complexity of training. If we have $N$ stocks and $F$ factors, the time complexity is $O(M N + N F^2 + F^3)$ where $M$ refers to runtime complexity of running the neural network on a single stock and the second and third terms are from the linear regression step. In other words, runtime complexity is linear in the number of stocks and cubic in the number of factors; often, in practice, we use far fewer factors than stocks. 

\subsection{Usage}\label{sec:usage}

\paragraph{Mean and Covariance} Unlike prior work that has worked on multivariate generative modeling (e.g., \cite{Wiese2021Multi,Sun2023GAN,Masi2023CorrelatedTimeSeries,tepelyan2023generative, Ramirez2023ML})), our model is able to compute the mean ($\bm\alpha + B^T \bm\mu_z$) and the covariance matrix without sampling ($\Sigma_x + B^T \Sigma_z B$) leading to significant speedup during inference.

\paragraph{One-day sampling} To sample, the stock embedder only needs to be run once per stock per day since, given $\alpha_{i,t}$, $\bm\beta_{i,t}$, $\sigma_{i,t}$ and $\nu_{i,t}$, the sampling is a simple function of samples from the prior distribution (\myref{Section}{sec:decoder}).

\paragraph{Multi-day sampling} Many variables prevent multi-day sampling such as the factors, volume, and options data since they change everyday but are not part of the generative model. If we were to assume company financials and industry are relatively static through time, we can sample multiple days by feeding back in the sampled stock returns into the stock embedder. In future work, since \myref{Section}{sec:ablation} finds the factors to be a useful set of features, we could experiment with replacing the factor indices by computing returns using portfolios of the stocks being modelled.

\section{Related Work}

\paragraph{Fama-French \citep{FamaFrench2015return}} The Fama-French factor model uses a predefined set of factor portfolios, similar to BDG. On the other hand, NeuralFactors learns factor exposures and factor portfolios are derived from these. 

\paragraph{BARRA \citep{barra1997}} Whereas NeuralFactors learns the factor exposures from the data, the BARRA style of factor modeling predefines the factor exposures through domain expertise using common factors such as value, momentum, and so forth.

\paragraph{PPCA} The original PCA approach of factor modeling lacks a generative and probabilistic interpretation; however probabilistic PCA (PPCA) can solve for this \citep{ghojogh2021factor}, which we compare against in \myref{Section}{sec:baseline}. PPCA is similar to NeuralFactors in that the factor exposures per company is inferred from the data itself. However, in PPCA, the exposures are only a function of the past stock returns;
our approach is able to have accurate time varying factor exposures through the exposures being a function of many other features.

\paragraph{Conditional Autoencoder \cite{Gu2021Auto}} While \citet{Gu2021Auto} used an auto\-encoder-style training, their work is more similar to BDG since, to handle the changing universe, \citet{Gu2021Auto} used a predefined set of latent factors. One difference between the two approaches is that the mean and other moments are nonlinear in BDG. Further, \citet{Gu2021Auto} lacks a generative and probabilistic interpretation.

\paragraph{Baseline Deep Generator (BDG) \citep{tepelyan2023generative}} This work did not make the linear assumption; however, it made strong assumptions about the sources of correlation.
Another difference is that in our model, once $\alpha_{i,t}$, $\bm\beta_{i,t}$, $\sigma_{i,t}$ and $\nu_{i,t}$ are computed per stock, we can sample without having to use a neural network unlike BDG. In other words, in NeuralFactors, the number of neural network forward passes is independent of the number of samples required. Further, due to the linear construction of the mean, the covariance matrix can be computed in closed form instead of through samples. 

\section{Results}\label{sec:eval}

Similar to \citet{tepelyan2023generative}, in our experiments, we use the point-in-time constituents of the S\&P 500 as our universe, a side effect of which is that the universe can change. 
We use daily data and split it into three pieces: the training set which contains data from the beginning of 1996 to the end of 2013, the validation set which contains data from the beginning of 2014 to the end of 2018, and the test set which contains data from the beginning of 2019 to the end of 2023. Note that we have an additional year of data in our test set. Importantly, all of our hyperparameter tuning was performed on the validation set without referring to the test set. 

For evaluation, we ablate our different modeling decisions, specifically the number of factors (\myref{Section}{sec:ablation_factors}), our choice of features (\myref{Section}{sec:ablation_features}), our choice of architecture and loss (\myref{Section}{sec:ablation_arch_loss}), our choice of lookback size (\myref{Section}{sec:ablation_lookback}), and our choice of the number of training years (\myref{Section}{sec:ablation_years}).

We then compare against some classical approaches, as well as prior work (\myref{Section}{sec:baseline}), showing that our generative approach is able to outperform all of them. For ease of comparison between models, we focus on two metrics \citep{tepelyan2023generative}:
\begin{align} 
\text{NLL}_{\text{joint}, t}  &= \frac{1}{N_{t+1}} \log p(\{r_{i,t+1}\}_{i=1}^{N_{t+1}} | \mathcal{F}_t)
\end{align}
\begin{align}
\text{NLL}_{\text{ind}, t} &= \sum_{i} \frac{1}{N_{t+1}}\log p(r_{i,t} | \mathcal{F}_t)
\end{align}
$\text{NLL}_{\text{joint}, t}$ measures the average negative log-likelihood of the conditional joint distribution across the universe at time $t$; $\text{NLL}_{\text{ind}}$ measures the average negative log-likelihood of the conditional univariate distributions  at time $t$. $\text{NLL}_{\text{joint}}$ denotes the average $\text{NLL}_{\text{joint}, t}$ across the evaluation period, and similarly for $\text{NLL}_{\text{ind}}$. 
\footnote{In order to reproduce our NLL numbers, we note that our returns are normalized by dividing by $0.02672357$, approximately the standard deviation of returns across our training period.} For our model, we approximate these integrals using 100 samples from the posterior distribution for $\text{NLL}_{\text{joint}, t}$ and 10K samples from the prior for $\text{NLL}_{\text{ind}, t}$.

Further, we compare our model against baselines in terms of VaR analysis (\myref{Section}{sec:var}), covariance forecasting (\myref{Section}{sec:covariance}), and portfolio optimization  (\myref{Section}{sec:portfolio}).

\subsection{Ablation Studies}\label{sec:ablation}

\begin{table*}[!bt]
    \centering
    \begin{tabular}{lc|lc|lc}
\toprule
\textit{Number of Factors} & 
& \textit{Features} & 
& \textit{Architecture}  & 
\\
\hline
64 Factors & $\mathbf{0.3240}$
& All Features & $\mathbf{0.3240}$
& Attention  & $\mathbf{0.3240}$
\\
\hline
128 Factors & {0.3249} 
& w/o Options and Volume  & 0.3430 
& LSTM  & 0.3419 
\\
32 Factors & 0.3289  
& Stock Returns, Financials, Industry  & 0.3629 
& $\alpha_{i,t} = 0$ & 0.3264 
\\
16 Factors & 0.3434
& Stock Returns   & 0.4656 
&  Gaussian Prior & 0.3256 
\\ 
 8 Factors & 0.3660
& &
& Gaussian Decoder & 0.4336 
\\ 
\hline
\hline

 \textit{Loss} & 
& \textit{Lookback Size} &
& \textit{Training Years} & \\  
\hline
 IWAE k=20 & $\mathbf{0.3240}$
& 256 &  $\mathbf{0.3240}$
& Last 18 Years & $\mathbf{0.3240}$ \\
\hline
VAE (IWAE k=1) & 0.3262 
& 192 &  0.3284 
& Last 15 Years & $0.3262$ 
\\
&
& 128 &  0.3291 
& Last 10 Years & $0.3348$
\\
 &
& &
& Last 5 Years & $0.3591$
\\ 
\bottomrule
\end{tabular}
    \caption{Ablation results.}
    \label{tab:ablate}
\end{table*}

For our ablation study, we focus on the $\text{NLL}_{\text{joint}}$ on the validation set since hyperparameter tuning including model design were based on this metric. All ablation results can be found in \myref{Table}{tab:ablate}.

\subsubsection{Choice of Number of Factors }\label{sec:ablation_factors}
For the number of factors, we tested powers of two from 1 to 128 and found that a total of 64 factors performs the best. In \myref{Table}{tab:ablate}, we see that choices of 32 and 128 factors perform worse.

\subsubsection{Choice of Features}\label{sec:ablation_features}
Our final model used stock returns, factor returns, company financials, industry, volume features, and option features. In \myref{Table}{tab:ablate}, we see that ablating each of these features reduces the overall performance of our model.

\subsubsection{Choice of Architecture and Loss}\label{sec:ablation_arch_loss}
Our final model used transformer blocks \citep{vaswani2017attention} for the sequence model; in \myref{Table}{tab:ablate}, we see that LSTM performs worse. 
Further, we find that our model performs better by allowing it to learn alpha ($\alpha_{i,t}$). In our architecture, we use Student's T distribution instead of a Gaussian which is more common in PPCA; here, we see that replacing either Student's T with a Gaussian hurts performance. Finally, we see that using the importance-weighted loss during training (we used $k=20$) leads to better performance over $k=1$.

\subsubsection{Choice of Lookback}\label{sec:ablation_lookback}
Our final model used a lookback size of 256, and, in \myref{Table}{tab:ablate}, we see monotonically decreasing performance with shorter lookback sizes. 

\subsubsection{Choice of Amount of Training Data}\label{sec:ablation_years} Prior work \citep{GUTIERREZ2019134,Pena2024CTGANFin} has claimed that the optimal length of data to train a generative model on is between three and five years.
In \myref{Table}{tab:ablate}, we see that the performance improved monotonically with more training data. In other words, the belief that non-stationarity requires us to use fewer years of training data appears to be incorrect and that more data consistently improves performance.

\subsection{Baselines}\label{sec:baseline}

For our baselines (\myref{Table}{tab:baseline_nll}, \ref{tab:baseline_cov}, \ref{tab:baseline_ce}, \ref{tab:baseline_sharpe}), we reproduced BDG \citep{tepelyan2023generative} and added additional features, namely industry, company financials, volume features, and options features; we refer to this model as BDG$^{*}$ in our tables. We refer to the results of the original BDG model as BDG$^{**}$. For this model, we reproduce the numbers reported in the original paper; we did not rerun these experiments since full evaluation of BDG takes 2000 GPU-hours whereas NeuralFactors takes approximately 24 GPU-hours to train and 1 GPU-hour to evaluate. 

Since we found additional features to help over the original set of features used in BDG (namely stock returns and factor indices) as well as using attention instead of LSTMs, we present the results of NeuralFactors with the original set of features (labeled ``Base Features'') and NeuralFactors using LSTM. 
Our results are organized by models trained with ``Additional Features'' (referring to models trained on the Base Features as well as Company Financials, Industry, Volume Features, and Options Features), ``Base Features'' (referring to models trained on Stock Returns and Factor Indices), and ``Baselines''. The best result is in \textbf{bold}, and the second best is \underline{underlined}.

For classical baselines, we include PPCA with 12 factors (found through hyperparameter tuning on the validation set) where we use a Student's T distribution for the decoder. We hypothesize the optimal number of factors for NeuralFactors is higher than PPCA because NeuralFactors is better able to distinguish factors from noise since the statistical strength of the model is improved through learning from features. Further, we include GARCH \citep{Engle1982ARCH} to compare performance on marginal distributions.

\subsubsection{Negative Log-Likelihoods}\label{sec:nll}

\begin{table}[!t]
    \centering
    \begin{tabular}{lcccc}
\toprule
     & \multicolumn{4}{c}{NLL ($\downarrow$) }
\\
 \cmidrule(lr){2-5} 
 
 & \multicolumn{2}{c}{Val}
 & \multicolumn{2}{c}{Test}

\\
 \cmidrule(lr){2-3} 
 \cmidrule(lr){4-5}
 & {Ind} & {Joint}
 & {Ind} & {Joint}
\\
\midrule
\textit{Additional Features} \\
NeuralFactors-Attention  
& ${0.747}$  & $\mathbf{0.324}$ 
& $\underline{1.029}$  & $\mathbf{0.556}$ 
 \\
NeuralFactors-LSTM  
& ${0.759}$  & $\underline{0.342}$ 
& ${1.041}$  & ${0.581}$
 \\
BDG$^{*}$ \citep{tepelyan2023generative}
& $\mathbf{0.726}$  & ${0.388}$ 
& $\mathbf{1.013}$  & ${0.620}$ 
\\
\midrule
\textit{Base Features} \\
NeuralFactors-Attention  
& ${0.760}$  & ${0.362}$ 
& ${1.044}$  & ${0.602}$ 
\\
NeuralFactors-LSTM  
& ${0.766}$  & ${0.378}$ 
& ${1.066}$  & ${0.619}$ 
 \\
BDG$^{**}$  \citep{tepelyan2023generative}
& $\underline{0.728}$ & $0.393$ 
& \na  & \na  
 \\
\midrule
\textit{Baselines} \\
PPCA (12 Factors) 
& ${1.016}$  & ${0.441}$ 
& ${1.326}$ & ${0.664}$
 \\
GARCH Skew Student 
& $0.774$  & \na 
& $1.053$  & \na 
 \\
\bottomrule
\end{tabular}
    \caption{Comparison against baseline models in terms of NLL (\myref{Section}{sec:nll})}
    \label{tab:baseline_nll}
\end{table}

In terms of $NLL_{joint}$, we can see in \myref{Table}{tab:baseline_nll} that, even with LSTM and the base features, we outperform BDG, implying that the source of improvements is not only the architectural improvements and features improvements, but also the new modeling methodology. Further, we observe that NeuralFactors is able to more effectively utilize additional features over BDG. In \myref{Table}{tab:ablate}, NeuralFactors improves by 0.0378 in the validation set by including the additional features (from 0.3618 to 0.3240), as opposed to BDG, which only improves by only 0.0051 (from 0.3875 to 0.3932).

We observe that PPCA performs similarly to NeuralFactors with only Stock Returns (\myref{Table}{tab:ablate}), which might be because both are functions of only the past stock returns.

We see that our model is able to outperform the GARCH and PPCA in terms of $NLL_{ind}$; however, BDG performs best on this metric. 

\subsubsection{Covariance Forecasting}\label{sec:covariance}

\begin{table}[!b]
    \centering
    \begin{tabular}{lcccccccccccccccccccc}
\toprule
     & \multicolumn{4}{c}{Covariance  ($\downarrow$)}
\\
 \cmidrule(lr){2-5} 
 
 & \multicolumn{2}{c}{Val}
 & \multicolumn{2}{c}{Test}
\\
 \cmidrule(lr){2-3} 
 \cmidrule(lr){4-5}
 & {MSE} & {Box's M}
 & {MSE} & {Box's M}
\\
\midrule
\textit{Additional Features} \\
NeuralFactors-Attention  
& $\mathbf{0.181}$ & $\mathbf{1.756}$ 
& ${0.282}$ & $\mathbf{2.226}$ 
\\
NeuralFactors-LSTM  
& $\underline{0.192}$ & $\underline{1.86}$ 
& ${0.369}$ & ${2.37}$ 
\\
BDG$^{*}$  \citep{tepelyan2023generative}
& ${0.225}$ & ${2.261}$ 
& ${0.332}$ & ${2.720}$ 
\\
\midrule
\textit{Base Features} \\
NeuralFactors-Attention  
& ${0.198}$ & ${1.939}$ 
& $\mathbf{0.246}$ & ${2.383}$ 
\\
NeuralFactors-LSTM  
& ${0.356}$ & ${2.358}$ 
& ${0.580}$ & ${2.917}$ 
\\
\midrule
\textit{Baselines} \\
PPCA (12 Factors) 
& ${0.378}$ & ${2.367}$ 
& ${0.881}$ & ${3.476}$ 
\\
\bottomrule
\end{tabular}
    \caption{Comparison against baseline models in terms of covariance forecasting (\myref{Section}{sec:covariance}).}
    \label{tab:baseline_cov}
\end{table}

To evaluate the quality of our covariance forecasts, we focus on the subset of stocks that are in the S\&P500 from the beginning of 2014 to the end of 2023. We are left with $s=324$ stocks. Using the mean and covariance forecasts, we whiten the observed next-day returns (${r}^{rot}_{t+1} = \Sigma_{t}^{-1/2} (\mathbf{r}_{t+1} - \bm\alpha_t)$); we compute the mean squared error (MSE) between the covariance of ${r}^{rot}_{t+1}$ and an identity matrix and compute Box's M test statistic \citep{box1949mtest}. In \myref{Table}{tab:baseline_cov}, we see that NeuralFactors consistently outperforms all other methods in terms of these two metrics. We see later in \myref{Section}{sec:portfolio} that, since classical portfolio optimization \citep{Markovitz1952} uses the mean and covariance, the investment performance of NeuralFactors outperforms the other models.

\subsubsection{Risk Analysis (VaR)}\label{sec:var}

\begin{table}[!bt]
    \centering
    \begin{tabular}{lcccccccccccccccccccc}
\toprule
     & \multicolumn{4}{c}{Calibration Error  ($\downarrow$)}
\\
 \cmidrule(lr){2-5} 
 
 & \multicolumn{2}{c}{Val}
 & \multicolumn{2}{c}{Test}

\\
 \cmidrule(lr){2-3} 
 \cmidrule(lr){4-5}
 & {Uni.} & {Port.}
 & {Uni.} & {Port.}
\\
\midrule
\textit{Additional Features} \\
NeuralFactors-Attention  
& ${0.110}$ & ${0.360}$ 
& $\underline{0.042}$ & ${0.074}$ 
\\
NeuralFactors-LSTM  
& ${0.136}$ & ${0.342}$ 
& ${0.048}$ & ${0.063}$ 
\\
BDG$^{*}$   \citep{tepelyan2023generative}
& ${0.092}$ & ${0.078}$ 
& ${0.086}$ & ${0.107}$ 
\\
\midrule
\textit{Base Features} \\
NeuralFactors-Attention  
& ${0.142}$ & ${0.395}$ 
& ${0.075}$ & ${0.133}$ 
\\
NeuralFactors-LSTM  
& ${0.105}$ & ${0.377}$ 
& $\mathbf{0.032}$ & $\underline{0.038}$ 
\\
BDG$^{**}$  \citep{tepelyan2023generative}
& $\mathbf{0.036}$ & $\mathbf{0.010}$ 
& \na  & \na  
\\
\midrule
\textit{Baselines} \\
PPCA (12 Factors) 
& ${0.810}$ & ${0.818}$ 
& ${0.741}$ & ${0.433}$ 
\\
GARCH Skew Student 
& $\underline{0.073}$ & $\underline{0.063}$ 
& ${0.049}$ & $\mathbf{0.005}$ 
\\
\bottomrule
\end{tabular}
    \caption{Comparison against baseline models in terms of VaR analysis (\myref{Section}{sec:var}). ``Uni.'' refers to taking a weighted average of calibration error per stock; ``Port.'' refers to the calibration error of an equal-weighted portfolio.  
    }
    \label{tab:baseline_ce}
\end{table}

One of the main values of having distributional estimates is in getting an estimate of the uncertainty, e.g., what is the probability the quantity is larger than zero. To evaluate the quality of our model in terms of risk, we use calibration error. 

\citet{kuleshov2018calibration} introduced calibration error as a metric to quantitatively measure how well the quantiles are aligned:
\begin{equation}
\begin{aligned}
 \hat{p}_j = {\abs{\{y_n  | \mathcal{F}_{x_n}(y_n) < p_j, n=1,\dots, N \}}}\ /\ {N} \\ 
 \text{cal}(y_1,\dots,y_N) = \sum_{j=1}^{M} (p_j - \hat{p}_j)^2
\end{aligned}\label{eqn:calib_err}
\end{equation}
where $\mathcal{F}_{x_n}$ is the predicted CDF function given $x_n$, $\hat{p}_j$ is the fraction of the data where the model CDF is less than $p$ and $M$ is the number of quantiles that are evaluated. In this work, we set this to $100$ evenly-spaced quantiles. This metric is zero when the fraction of the data where the model CDF is less than $p$ is $p$.

In \myref{Table}{tab:baseline_ce}, we compute the average calibration error across stocks weighed by the number of days the stock is in the S\&P 500 in the corresponding period (``Uni.'') and the calibration error of an equal-weighted portfolio comprised of the point-in-time stocks in the S\&P 500 (``Port.''). In the validation set, NeuralFactors performs better only against PPCA and worse compared to all the other models; however, in the test set, NeuralFactors performs the best in the Uni. calibration error and second best on the portfolio calibration error. Given that the covariance forecasts made by NeuralFactors is better than GARCH and BDG (\myref{Section}{sec:covariance}), the performance in terms of calibration error implies the tails have not been modeled completely; we leave it to future work to close this gap in performance.

\subsubsection{Portfolio Optimization}\label{sec:portfolio}

\begin{table*}[!bt]
    \centering
    \begin{tabular}{lcccccccccccccccccccc}
\toprule
     & \multicolumn{8}{c}{Sharpe  ($\uparrow$)}
\\
 \cmidrule(lr){2-9} 
 
 & \multicolumn{4}{c}{Val}
 & \multicolumn{4}{c}{Test}

\\
 \cmidrule(lr){2-5} 
 \cmidrule(lr){6-9}
 & {L} & {L/S} & {L Lev. 1} & {L/S Lev. 1}
  & {L} & {L/S} & {L Lev. 1} & {L/S Lev. 1}
\\
\midrule
\textit{Additional Features} \\
NeuralFactors-Attention  
&  $\mathbf{1.87}$ & $\underline{3.68}$ & $\mathbf{1.72}$ & ${1.35}$ 
& $\underline{1.2}$ & $\mathbf{2.54}$ & $\underline{1.04}$ & ${1.3}$ 
\\
NeuralFactors-LSTM  
& ${1.52}$ & ${3.36}$ & ${1.59}$ & $\mathbf{1.51}$
& $\mathbf{1.32}$ & $\underline{2.51}$ & $\mathbf{1.32}$ & $\mathbf{1.66}$ 
\\
BDG$^{*}$   \citep{tepelyan2023generative}
& $1.61$ & $3.47$ & $0.99$ & $\underline{1.42}$ 
& $0.84$ & $2.33$ & $0.75$ & $0.84$ 
\\
\midrule
\textit{Base Features} \\
NeuralFactors-Attention  
& $\underline{1.72}$ & $\mathbf{3.80}$ & ${1.54}$ & ${1.36}$
& ${0.46}$ & ${1.79}$ & ${0.55}$ & ${1.38}$
\\
NeuralFactors-LSTM  
& ${1.48}$ & ${2.39}$ & ${1.42}$ & ${1.06}$
& ${0.6}$ & ${1.98}$ & ${0.61}$ & $\underline{1.45}$ 
\\
\midrule
\textit{Baselines} \\
PPCA (12 Factors) 
& ${0.43}$ & ${-0.07}$ & ${0.46}$ & ${0.44}$ 
& ${0.56}$ & ${0.02}$ & ${0.56}$ & ${0.34}$  
\\
\bottomrule
\end{tabular}
    \caption{Comparison against baseline models in terms of portfolio optimization (\myref{Section}{sec:portfolio}). 
    ``L'' refers to Long-Only strategy, ``L/S'' refers to a Long-Short strategy, and ``Lev. 1'' refers to $L=1$ in \myref{Equation}{eqn:port_opt}.
    }
    \label{tab:baseline_sharpe}
\end{table*}

\begin{figure}[!b]
    \centerline{\includegraphics[width=0.95\linewidth]{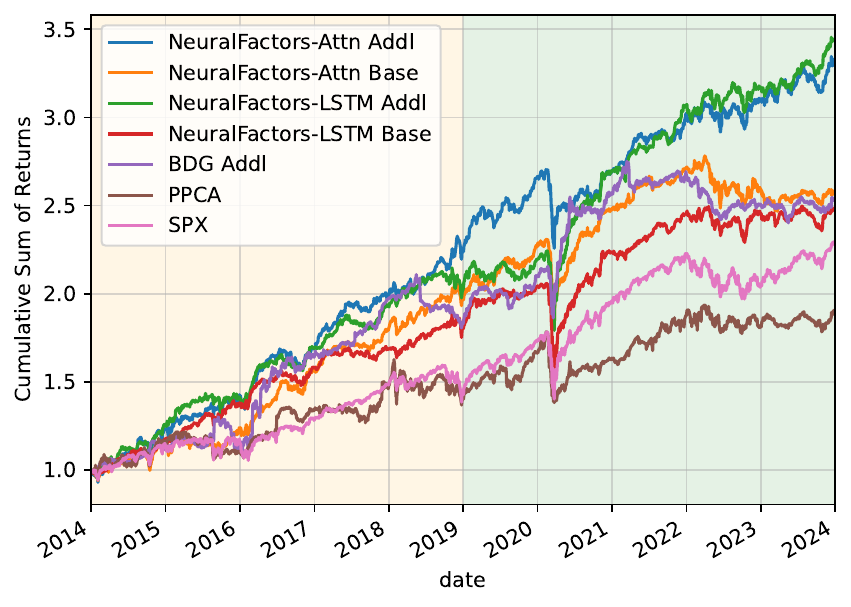}}
    \caption{Comparison of returns of long-only $L=1$ portfolios. To make all of our final
    results comparable in scale, we lever the returns to match the volatility of the S\&P 500.}
\label{fig:port_returns}
\end{figure}

\begin{figure}[!tb]
    \centerline{\includegraphics[width=0.95\linewidth]{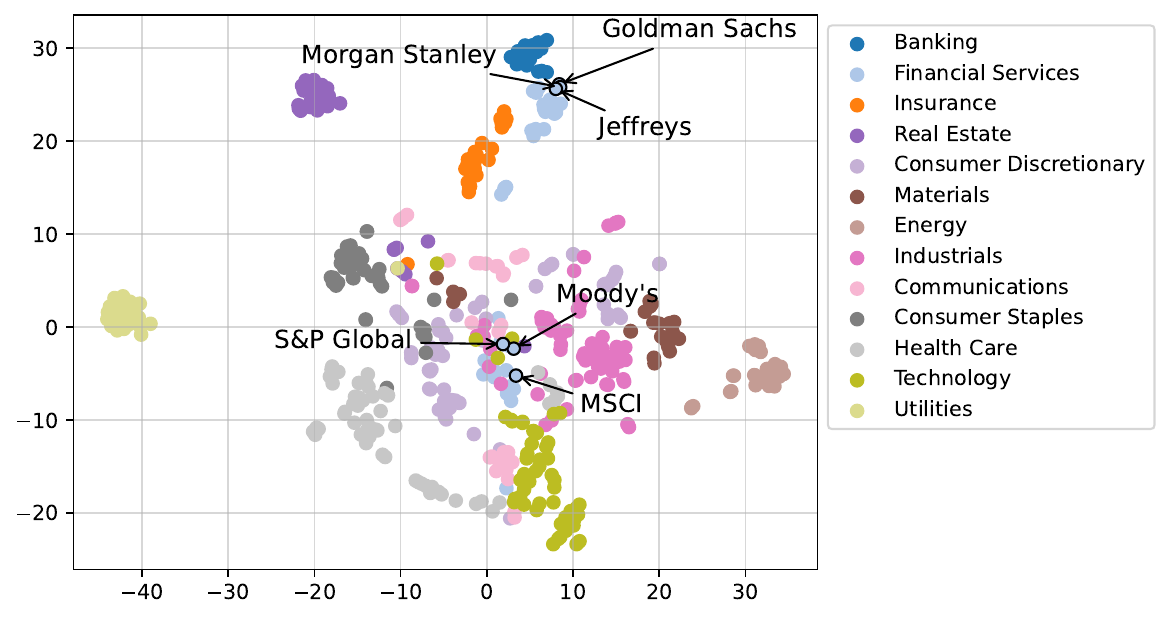}}
    \caption{TSNE embedding of $\bm\beta_{i,t}$ for $t=$ 03 Jan 2019.}
\label{fig:tsne}
\end{figure}

A common use case of covariance estimates is portfolio optimization \citep{Markovitz1952}. For simplicity, we focus on mean-variance optimization:
\begin{equation}\label{eqn:port_opt}
 \argmax_{\bm{w} \in \RR^{N},\ \norm{w}_1 = L} \mathbb{E}_{\bm{r}_{T+1}}\left[ \bm{w}^T \bm{r}_{T+1} \right] - \frac{\lambda}{2} \mathbb{V}_{\bm{r}_{T+1}}\left[ \bm{w}^T \bm{r}_{T+1} \right] 
\end{equation}

We experiment with four configurations, long only with $L=\infty$, long only with $L=1$,  long-short with $L=\infty$, and  long-short with $L=1$. In \myref{Table}{tab:baseline_sharpe}, we see, for the validation period, that NeuralFactors-Attention with Additional Features is one of the top two for all but the Long/Short $L=1$, for which it does worse than BDG and NeuralFactors-LSTM with Additional Features. Similarly for the test period, NeuralFactors-Attention with Additional Features is one of the top two for all but the Long/Short $L=1$ for which it does worse than NeuralFactors-LSTM with Additional Features and  NeuralFactors-LSTM with Base Features. We note that across all the different configurations portfolio optimization configurations, the model with the highest $NLL_{joint}$ performs the best most often.

\subsection{Qualitative Analysis}\label{sec:tsne}

In \myref{Figure}{fig:tsne}, we show that $\beta_{i,t}$ can be used as an embedding stocks. Specifically, we embed $\beta_{i,t}$ into two dimensions using TSNE \citep{Maaten2008TSNE} and see that the data naturally clusters around sectors. Further, we note that even though there are two clusters for Financial Services companies, each cluster contains companies that are considered peers, namely Goldman Sachs, Morgan Stanley, and Jeffreys are in one cluster and MSCI, Moody's and S\&P Global are in the other.

\section{Conclusion}

In this paper, we introduced a novel modeling methodology for applying machine learning to factor analysis which we call NeuralFactors. The training methodology is based on variational inference; because of this connection, we leave it to future work to explore combining NeuralFactors with VAE-based imputation \citep{mattei2019miwae}. 

We found attention significantly outperforms LSTMs; we leave it to future work to explore other sequence models such as S4 \citep{gu2022s4} and xLSTM \citep{beck2024xlstm}. We observed the learned factor exposures ($\beta_{i,t}$) cluster according to similarity; we leave it to future work to explore using interpretability techniques such as SHAP \citep{Lundberg17SHAP} in order to understand the learned factors.

Notable, NeuralFactors, when applied to equity returns,  outperforms probabilistic PCA (PPCA) and BDG \citep{tepelyan2023generative}, a previous approach to generative modeling for equities in terms of negative log-likelihoods, covariance forecasting, and portfolio optimization.
Given that NeuralFactors is able to effectively utilize features that are atypical in factor modeling, we hope to explore in the future the usage of news-based factors, supply chain factors, and other alternative datasets. 

Finally, we note that NeuralFactors is a generic approach to factor analysis and can be extended to other domains such as modeling equities in other market, modeling yields, and even non-financial domains in which factor analysis is used.

\bibliographystyle{ACM-Reference-Format}
\bibliography{sample}

\newpage
\appendix

\end{document}

%% file: final_arch.tex
\begin{figure*}[!bt]
  \begin{center}
\scalebox{1.0}{
  \begin{tikzpicture}[
    every neuron/.style={
      circle,
      minimum size=0.3cm,
      very thick
    },
    every data/.style={
      rectangle,
      minimum size=0.4cm,
      thick
    },
  ]
  
    \node [align=center,every neuron/.try, data 1/.try, minimum width=0.3cm] (stock-n)  at ($ (2.5, 0) $) {$\text{Stock}_{N, t}$};

    \node [align=center,every neuron/.try, data 1/.try, minimum width=0.3cm] (input-3)  at ($ (stock-n) + (0,1.5) $) {$\mathbf{\vdots}$};
    \node [align=center,every neuron/.try, data 1/.try, minimum width=0.3cm] (stock-1)  at ($ (input-3) + (0,1.5) $) {$\text{Stock}_{1,t}$};

    \node [align=center,every data/.try, data 1/.try, minimum width=0.3cm, draw] (stock-emb-1)  at ($ (stock-1) + (2.5, 0) $) {Neural Network};
    \node [align=center,every data/.try, data 1/.try, minimum width=0.3cm, draw] (stock-emb-n)  at ($ (stock-n) + (2.5, 0) $) {Neural Network};

    \draw [black,solid,->] ($(stock-1.east)$) -- ($(stock-emb-1.west)$);
    \draw [black,solid,->] ($(stock-n.east)$) -- ($(stock-emb-n.west)$);

    \node [align=center,every data/.try, data 1/.try, minimum width=0.3cm] (alpha-1)  at ($ (stock-emb-1) + (2.0, 0.75) $) {$\alpha_{1,t}$};
    \node [align=center,every data/.try, data 1/.try, minimum width=0.3cm] (beta-1)  at ($ (stock-emb-1) + (2.0, 0.25) $) {$\bm\beta_{1,t}$};
    \node [align=center,every data/.try, data 1/.try, minimum width=0.3cm] (sigma-1)  at ($ (stock-emb-1) + (2.0, -0.25) $) {$\sigma_{1,t}$};
    \node [align=center,every data/.try, data 1/.try, minimum width=0.3cm] (nu-1)  at ($ (stock-emb-1) + (2.0, -0.75) $) {$\nu_{1,t}$};

    \draw [black,solid,->] ($(stock-emb-1.east)$) -- ($(alpha-1.west)$;);
    \draw [black,solid,->] ($(stock-emb-1.east)$) -- ($(beta-1.west)$;);
    \draw [black,solid,->] ($(stock-emb-1.east)$) -- ($(sigma-1.west)$;);
    \draw [black,solid,->] ($(stock-emb-1.east)$) -- ($(nu-1.west)$;);
    
    \node [align=center,every data/.try, data 1/.try, minimum width=0.3cm] (alpha-n)  at ($ (stock-emb-n) + (2.0, 0.75) $) {$\alpha_{N,t}$};
    \node [align=center,every data/.try, data 1/.try, minimum width=0.3cm] (beta-n)  at ($ (stock-emb-n) + (2.0, 0.25) $) {$\bm\beta_{N,t}$};
    \node [align=center,every data/.try, data 1/.try, minimum width=0.3cm] (sigma-n)  at ($ (stock-emb-n) + (2.0, -0.25) $) {$\sigma_{N,t}$};
    \node [align=center,every data/.try, data 1/.try, minimum width=0.3cm] (nu-n)  at ($ (stock-emb-n) + (2.0, -0.75) $) {$\nu_{N,t}$};

    \draw [black,solid,->] ($(stock-emb-n.east)$) -- ($(alpha-n.west)$;);
    \draw [black,solid,->] ($(stock-emb-n.east)$) -- ($(beta-n.west)$;);
    \draw [black,solid,->] ($(stock-emb-n.east)$) -- ($(sigma-n.west)$;);
    \draw [black,solid,->] ($(stock-emb-n.east)$) -- ($(nu-n.west)$;);
    
    \node [align=center,every data/.try, data 1/.try, minimum width=0.3cm] (q)  at ($ (nu-n) + (3.0, -1.75) $) {$\mathbf{z_{t+1}} \sim \mathcal{N}\left( \Sigma_{z | B, t}\ \left(\Sigma_z^{-1} \bm\mu_z + B^T \Sigma_{x,t}^{-1} (\mathbf{r_{t+1}} - \bm{\alpha_{t}} ) \right)\ ,\ \Sigma_{z | B, t}\right)$ \\
    $\Sigma_{z | B, t} = \left(\Sigma_{z}^{-1} + B_t^T \Sigma_{x, t}^{-1} B_t\right)^{-1}$ \\ $p(z) = \mathbf{t}_{\bm\nu_z}(\bm\mu_z, \bm\sigma_z)$};

    \path [black,solid,->, opacity=0.5] ($(alpha-1.east)$) edge [bend left=20]  ($(q.north) + (0.1, 0)$);
    \path [black,solid,->, opacity=0.5] ($(beta-1.east)$) edge [bend left=20]  ($(q.north)+ (0.1, 0)$);
    \path [black,solid,->, opacity=0.5] ($(sigma-1.east)$) edge [bend left=20]  ($(q.north)+ (0.1, 0)$);
    \path [black,solid,->, opacity=0.5] ($(nu-1.east)$) edge [bend left=20]  ($(q.north)+ (0.1, 0)$);

    \path [black,solid,->, opacity=0.5] ($(alpha-n.east)$) edge [bend left=20]  ($(q.north)- (0.11, 0)$);
    \path [black,solid,->, opacity=0.5] ($(beta-n.east)$) edge [bend left=20]  ($(q.north)- (0.1, 0)$);
    \path [black,solid,->, opacity=0.5] ($(sigma-n.east)$) edge [bend left=20]  ($(q.north)- (0.095, 0)$);
    \path [black,solid,->, opacity=0.5] ($(nu-n.east)$) edge [bend left=20]  ($(q.north)- (0.09, 0)$);

    \node [align=left,every data/.try, data 1/.try, minimum width=0.3cm] (r-1)  at ($ (stock-emb-1) + (10.0, 0.0) $) {$p\left(r_{1, t+1}\ |\ t_{\nu_{1,t}}(\alpha_{1,t}+ \bm\beta_{1,t}^T\mathbf{z_{t+1}}, \sigma_{1, t})\right)$};

    \path [black,solid,->, opacity=0.5] ($(alpha-1.east)$) edge [bend left=20]  ($(r-1.north)$);
    \path [black,solid,->, opacity=0.5] ($(beta-1.east)$) edge [bend left=20]  ($(r-1.north)$);
    \path [black,solid,->, opacity=0.5] ($(sigma-1.east)$) edge [bend right=20]  ($(r-1.south)$);
    \path [black,solid,->, opacity=0.5] ($(nu-1.east)$) edge [bend right=20]  ($(r-1.south)$);
    
    \node [align=left,every data/.try, data 1/.try, minimum width=0.3cm] (r-n)  at ($ (stock-emb-n) + (10.2, 0.0) $) {$p\left(r_{N, t+1}\ |\ t_{\nu_{N,t}}(\alpha_{N,t}+ \bm\beta_{N,t}^T\mathbf{z_{t+1}}, \sigma_{N, t})\right)$};

    \path [black,solid,->, opacity=0.5] ($(alpha-n.east)$) edge [bend left=20]  ($(r-n.north)$);
    \path [black,solid,->, opacity=0.5] ($(beta-n.east)$) edge [bend left=20]  ($(r-n.north)$);
    \path [black,solid,->, opacity=0.5] ($(sigma-n.east)$) edge [bend right=20]  ($(r-n.south)$);
    \path [black,solid,->, opacity=0.5] ($(nu-n.east)$) edge [bend right=20]  ($(r-n.south)$);

      \path [black,solid,->] ($(q.north) - (3.0,0)$) edge [bend left=20]  ($(r-n.west)$);
      \path [black,solid,->] ($(q.north) - (3.0,0)$) edge [bend left=20]  ($(r-1.west)$);

      \draw[red,thick,dotted] ($(alpha-1.north east)+(0.25,0.2)$)  rectangle ($(stock-n.south west)-(0.35,0.5)$);
      \node [align=center,data 1/.try, minimum width=1.5cm] (stock-emb-desc)  at ($(stock-1) - (-0.3, -0.8)$) {\textit{Stock Embedder}};

      \draw[red,thick,dotted] ($(q.north east)+(0.25,0.15)$)  rectangle ($(q.south west)-(0.35,0.)$);
      \node [align=center,data 1/.try, minimum width=1.5cm] (q-desc)  at ($(q.south) + (3.2, 0.25)$) {\textit{Encoder}};

      \draw[red,thick,dotted] ($(alpha-1.north east)+(10.45,0.2)$)  rectangle ($(stock-n.south west)+(10.9-0.35,-0.5)$);
      \node [align=center,data 1/.try, minimum width=1.5cm] (des-desc)  at ($(stock-1) + (14.54, 0.8)$) {\textit{Decoder}};
    
  \end{tikzpicture}
  }
  \end{center}
  \caption{We show a high-level diagram of our final model architecture. $\mathbf{r_{i, t}}$  denotes the returns of security $i$ at time $t$. Note that the ``Neural Network'' is the same across all stocks.}\label{fig:final_arch}
  
  \end{figure*}